\documentstyle[12pt,epsfig]{article}

\begin{document}

\vspace{1cm}
\begin{flushright} 
{\bf DFUB 2001-11}\\
Bologna, \today
\end{flushright}
\vspace{1 cm}
\begin{center}

{\large {\bf SEARCH FOR RADIATIVE DECAYS OF SOLAR NEUTRINOS\\
 DURING A SOLAR ECLIPSE}

\vspace{0.5 cm}

G. Giacomelli$^1$ and V. Popa$^{1,2}$}

\vspace{0.5 cm}

1.{\em Dipartimento di Fisica, Universit\`a di Bologna and INFN,
 Bologna, 40127, Italy}\\
  E-mail: giacomelli@bo.infn.it

2. {\em Bucharest - M\u{a}gurele, 76900, Romania}\\
 E-mail: popa@bo.infn.it, vpopa@venus.nipne.ro

\vspace{5mm}

Invited paper at NO-VE, Int. Workshop on Neutrino Oscillations in Venice, \\
Venice, Italy, July 24-26, 2001.

\vspace{2cm}

Abstract

\end{center}

{\small A search for possible radiative decays of solar neutrinos with 
emission of photons in the visible range may be performed during total 
solar eclipses. We discuss some results obtained from the digitized images
recorded during the August 11, 1999 total solar eclipse
in Romania, and report on the
observations made in June 21, 2001, in Zambia.}
   
\normalsize\baselineskip=15pt

\section{Introduction}

It is a general opinion that most probably neutrinos have
non--zero masses. This belief is based primarely on the evidence/indication
for neutrino oscillations from data on solar and atmospheric
neutrinos.

Neutrino oscillations are possible if the flavour eigenstates ( $\nu_e$, 
$\nu_\mu$, $\nu_\tau$) are
not pure mass eigenstates ($\nu_1$, $\nu_2$, $\nu_3$), e.g.:
\begin{equation}
        |\nu_e> = |\nu_1> \cos \theta + |\nu_2> \sin \theta
\end{equation}
where  $\theta$ is the mixing angle and $m_{\nu_2} > m_{\nu_1}$.

Since few years there is evidence that the number of
solar neutrinos arriving on Earth is considerably smaller than
what is expected on the basis of the ``Standard Solar Model'' and of
the ``Standard Model'' of particle physics, where neutrinos are
massless (see e.g.\cite{Bahcall}).
One possible explanation of these experimental results involves
neutrino oscillations, either in vacuum with $\Delta m^2_{sun} =
m_{\nu_2}^2-m_{\nu_1}^2 \sim 10^{-10}$ eV$^2$
(as originally discussed in refs.\cite{Gribov,Pontecorvo})
or resonant matter oscillations 
$\Delta m^2_{MSW} \sim 10^{-5}$ eV$^2$ \cite{Wolfenstein,Mikheyev}.
 Recent results from the Super-Kamiokande,\cite{SK} MACRO\cite{macro95,MACRO}
and Soudan2 \cite{Soudan2}
experiments on atmospheric neutrinos strongly support the hypothesis of
atmospheric neutrino oscillations, in particular $\nu_{\mu} \rightarrow
\nu_{\tau}$, with large mixing
($\sin^2 2\theta > 0.8$) and $\Delta m^2_{atm} \simeq 2.5 \times 10^{-3}$
eV$^2$.

Another indication in favor of
neutrino oscillations with a third energy scale
$\Delta m^2_{LSND} \simeq 1$ eV$^2$
was reported in ref.\cite{LSND}.

Very recently, the SNO experiment reported evidence of a ``non-electron
flavor active neutrino component in the solar flux".\cite{sno} This implies
the observation of solar active $^8$B neutrinos in close agreement with the
predictions of solar models, and thus it is an important result in favor of 
solar neutrino oscillations. 

From the ensamble of data concerning solar and atmospheric neutrinos and 
considering also the beta spectrum of tritium,\cite{trit} the sum of the 
masses of active neutrinos is estimated to be between 0.05 and 8.4 eV.\cite{sno}

The above observations appear to be the first indications
for new physics beyond the ``Standard Model";  any
model that generates neutrino masses must contain a natural mechanism
that explains their values and the relation to the masses of
their corresponding charged leptons.
Different scenarios have been
proposed to explain all the observations, including the results with
neutrinos from reactors and accelerators.\cite{Boehm,Mezzetto,GG}

If neutrinos do have masses, then the heavier neutrinos could
decay into the lighter ones. For neutrinos with masses of few eV the only
decay modes kinematically allowed are radiative decays of the type
$\nu_i \rightarrow \nu_j + \gamma$ (where lepton flavor would be violated).
Such decay processes 
in astrophysics were first suggested by Masiero and Sciama.\cite{sciama}

Upper bounds on the lifetimes of such decays are
based on the astrophysical non--observation of the final state $\gamma$ rays.
Limits were obtained from measurements of $X$ and $\gamma$ ray fluxes
from the Sun\cite{Raffelt} and SN 1987A.\cite{Frieman,Chupp}

In the case of neutrinos with nearly degenerated masses,
of the order of the eV, the emitted photon can
be in the visible or ultraviolet bands.\cite{Bouchez,Oberauer,Birnbaum,frere}
A first tentative to detect such photons, using the Sun as a
source, was made during the total solar eclipse of
October 24, 1995.\cite{Birnbaum}

Direct visible photons from the Sun come at a rate of some $10^{17}$
cm$^{-2}$ s$^{-1}$; this makes a direct search for photons from
solar neutrino decays
impossible. To perform a measurement one must take advantage
of a total solar eclipse, which reduces by at least 8 orders of
magnitude the direct photon flux.
By looking with a telescope at the dark disk of the Moon during a solar eclipse 
one
can search for photons emitted by neutrinos decaying during their
380000 km flight path from the Moon to the Earth.

In this paper we describe the methodology employed for such a search,
give limits obtained from a preliminary measurement performed
during the total solar eclipse of 11 August, 1999, and we report on the
observations made in Zambia, the 21 June, 2001.

\section{Kinematics of radiative decays}

We assume the existence of a possible neutrino radiative decay,
$\nu_2 \rightarrow \nu_1 + \gamma$, where $m_{\nu_2} > m_{\nu_1}$;
 $\nu_1$, $\nu_2$ are neutrino mass eigenstates .

The energy of the emitted photon in the earth reference laboratory system is

\begin{equation}
        E_{lab}=E_{cm} \gamma_{\nu} \left( 1+ \beta_{\nu}
        \cos \theta^{\ast} \right),
\end{equation}
where $E_{\nu}$ and $\gamma_{\nu}=\frac{E_{\nu}}
{m_{\nu_2}}={\left( 1 - \beta^2_{\nu}
\right)}^{-\frac{1}{2}}$ are in the lab. frame;
 $\theta^{\ast}$ and $E_{cm}$ are
the photon emission angle with respect to the parent neutrino spin 
direction, and the energy of the emitted
photon in the decaying neutrino rest frame.

For radiative neutrino decays the general
expression for the angular distribution of the emitted photons in
the rest frame of the parent neutrino is

\begin{equation} \label{eqn:1}
        \frac{dN}{d\cos\theta^{\ast}}=\frac{1}{2} \left( 1-
        \alpha \cos\theta^{\ast} \right)
\end{equation}
where the $\alpha$ parameter is equal to -1, +1, for left--handed and
right--handed Dirac neutrinos, respectively; it is 0 for Majorana neutrinos.

In order to estimate the expected fraction  of photons 
produced in the visible range and
their maximum angle of emission  from radiative
solar neutrino decays we performed Monte Carlo
simulations for neutrino masses in the range $0.1-4$ eV and two
different sets of oscillation parameters
(large mixing: $\Delta m^2= 2 \times 10^{-4}$
eV$^2$ and $\sin^2 2\theta=0.71$; small mixing: $\Delta m^2= 6 \times 10^{-6}$
eV$^2$ and $\sin^2 2\theta=3.98 \times 10^{-3}$), coresponding to  the extremes
of the Super-Kamiokande  allowed regions for MSW oscillations in matter.

Fig: 1 is the result of $5 \times 10^8$ simulated
radiative decays, for each ($m_{\nu_1}$, $\Delta m^2$) set
of values, and assuming that neutrinos are left-handed.

\begin{figure}
\begin{center}
\vspace {-1.5 cm}
\mbox{\epsfysize=8 cm
      \epsffile{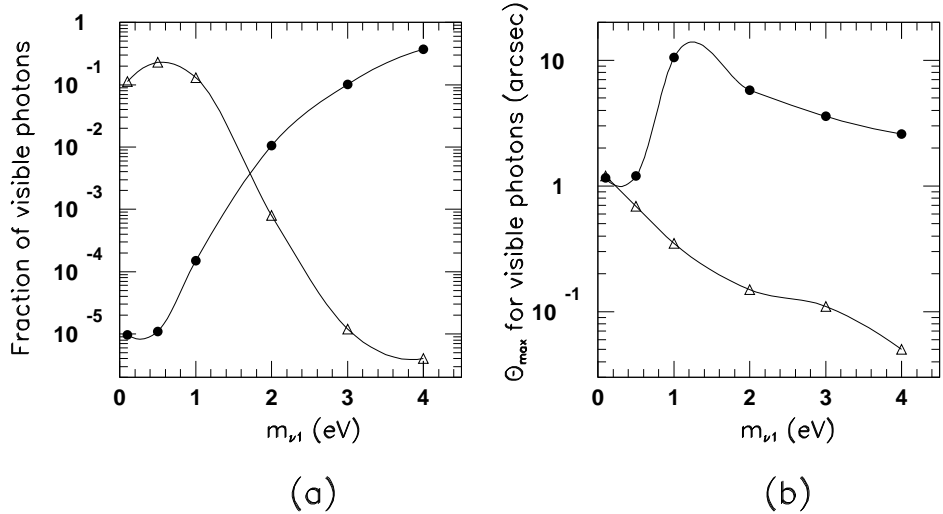}}
\end{center}

\vspace{-1.5cm}
\caption{\small
Summary of the Monte Carlo simulations of solar neutrino radiative
decays: (a) the fraction of decays leading to visible photons, and (b)
the maximum emission angle with respect to the Sun - Earth direction for the 
visible photons. 
The black points refer to the ($\Delta m^2= 2 \times 10^{-4}$
eV$^2$, $\sin^2 2\theta=0.71$) oscillation parameters, while the white
triangles refer to the ($\Delta m^2= 6 \times 10^{-6}$
eV$^2$,  $\sin^2 2\theta=3.98 \times 10^{-3}$) set. The curves are only drown 
to guide the eye.}
\end{figure}
 
\section{The August 11, 1999 total solar eclipse}

Two experiments were prepared for the observation of  solar neutrino 
radiative decay signatures, during the solar eclipse of August 11, 1999.

The first experiment used a small Cassagrain telescope mounted on an automatic 
pointing device;  the whole apparatus was carried on the rear seat of a 
MIG-29 supersonic fighter of the Romanian Air Forces, that was supposed to 
follow the totality of the eclipse in Romania. 

A second experiment, 
used a larger
Newtonian telescope, that was supposed
 to make observations from a site very close to the point 
of maximum eclipse, in the Par\^{a}ng Mountains, in South-Western Romania.
Both telescopes were equipped with CCD cameras, acquisition and 
pointing control and recording computers.
Combining the results of the two experiments, we expected an improvement 
of at least a factor of 20 \cite{iordania}
 compared to a previous experiment of the same 
kind.\cite{Birnbaum} 

Unfortunately, both experiments failed due to the extremeley 
unfavorable weather 
conditions that made the plane takeoff too hazardous, and the observation
from ground impossible.

By the courtesy of the V\^{a}lcea 1 Television in R\^{a}mnicu V\^{a}lcea, 
Romania, 
we obtained a S-VHS recording of the eclipse, that could be analyzed in a 
 way similar to what we intended to do with the CCD images. 
The film allowed the analysis in three colours (red, blue and green) and of 
their sum (white).
As no direct 
calibrations were possible, the numerical results obtained are not accurate, 
but could still offer a qualitative understanding on the phenomenology 
under study.

After digitising the film, we selected 2747 good quality frames, each 
352$\times$288 pixels$^2$. For each frame we extracted a 32$\times$32 
pixels$^2$ 
large square\footnote{The size of 32$\times$32 
pixels$^2$ is the largest dyadic (integer
power of 2) size 
of a square that fits inside the dark Moon disk.},
centered on the center of the dark disk of the Moon (which clearly
corresponded to the center of the Sun behind) and summed them in a unique 
``image". 

Repeating the same selection on a real full Moon image of the same
size, we could observe that the structures present in our composed picture
were produced by the moonscape seen in the light reflected by the Earth; 
 after suitable normalisations (in intensity and dynamics) we could remove these
structures from our data.

In order to make an indirect calibration of our Aquisition Digital Units
(ADU's) we used as reference the measurement of the luminosity distribution
of the solar corona, made during the same eclipse, by a group of astronomers
from the Pises Observatory, France.\cite{pises} Using their estimates and
the average brightness of the full Moon (0.34 lux, which, assuming an average
wavelength of 5500 $\AA$ corresponds to  $\simeq 1.4 \times 10^{11}$ 
photons cm$^{-2}$s$^{-1}$ at the earth surface) we could estimate the flux
of visible photons that would produce 1 ADU in our image.

Electron neutrinos are produced in termonuclear reactions in a small inner 
region of the Sun (about 0.6\% of the 
solar diameter corresponding to 12" as seen from the Earth). 
 The decay visible photons are emitted 
nearly along the direction of incidence of the parent neutrinos;  since
at our resolution the angular size coresponding to one pixel is about 
20", we expect that the decay signal would be present only in the central 
pixel of the image, or in its adiacent pixels.

  We are interested to search for  a 
small effect in the data; thus we applied a wavelet decomposition
(see e.g. ref.\cite{wave}) of 
the image, after the removal of the moon background.

The average 
(over the azimuthal angle) 
residual light fluxes corresponding to the fourth order residual 
in the wavelet decomposition, are presented in Fig: 2,
for the red, blue, green and white light. 
Note that the negative 
flux values at some distances from the center are an artefact due to the 
wavelet analysis of noisy signals, and indicate the degree of fluctuations in 
our measurement.  

\begin{figure}
\vspace{-1.5 cm}
\begin{center}
\mbox{\epsfysize=10 cm
      \epsffile{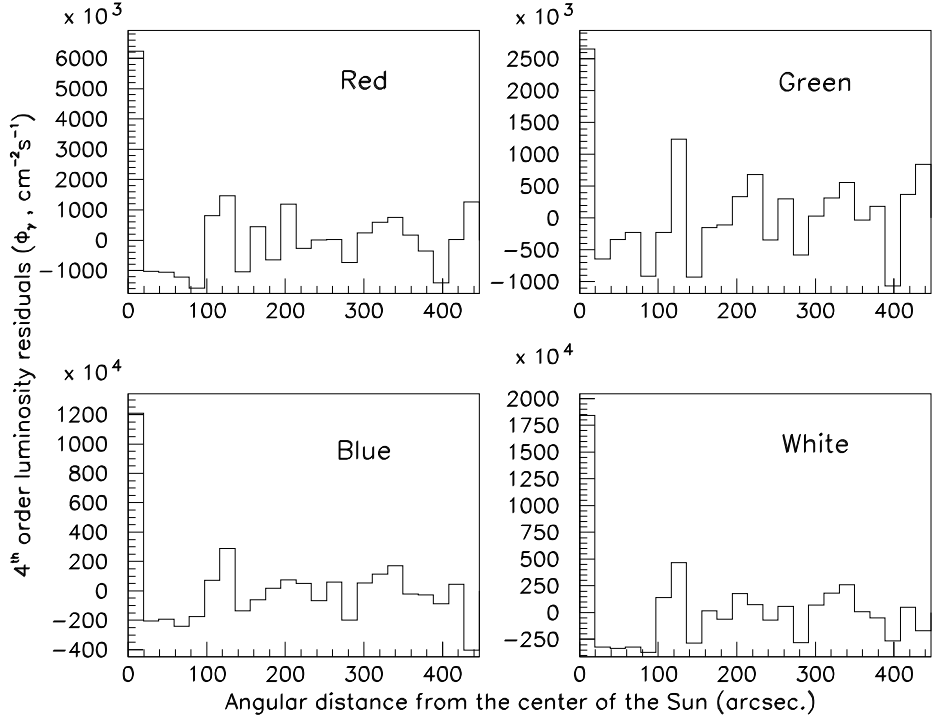}}
\end{center}

\vspace{-1 cm}
\caption{\small
Average residual light fluxes, after Moon substraction and wavelet
decomposition, in the red, green, blue channels and summed (white), as a 
function of the angular distance from the 
center of the Moon disk.}
\end{figure}

It should be mentioned that the videocamera produces the green channel by 
subtracting  the red and the
blue channels from the image obtained without filters (white); 
thus the green channel is somewhat artificial. 

After the 
digitization of the video recording, we reconstructed the white channel
as the sum of the blue, red and the fake green channels.
We checked the consistency of 
the peaks corresponding to the central pixels in 
the four color channels  (see Fig: 2)
by repeating the analysis considering other pixels of the image as ``central".
In  these cases we did not obtain the same  structures as in Fig: 2.

We tried  to explain 
the peaks at small angular distances 
 assuming that the 
central maxima are due to some diffraction pattern. In the particular case 
of a circular opaque object 
(in our case the Moon)
whose size is much larger than the wavelength of the
light emitted by the source (the Sun), 
but much smaller than the distance from the object to 
the source to the screen (detector),
 an ``unusual" diffraction effect known as ``the 
Poisson spot" occurs.\cite{lomel,som,rinard} It consists 
in a strong diffraction
maximum at the center of the shadow of the circular object; for 
point sources its intensity should be the same as if the circular obstacle 
was not there; for plane waves it should be a quarter of this. 
The Moon is not a perfect circular obstacle and the Sun is neither a point 
source of light, nor a source of plane waves, furthermore the apparent 
angular dimension of the solar corona is larger than that of the Moon, 
thus we are
not in the ideal case of the production of the Poisson spot; on the other 
hand, the distance conditions are fully fullfiled in the case of a total 
solar eclipse seen from Earth (the town of R\^{a}mnicu V\^{a}lcea was only 
few km from the point of maximum eclipse), so we could expect that such a 
diffraction effect is present in our data. 

In the diffraction hypothesis,
the intensity of the central diffraction maximum should not be dependent on 
the wavelength; thus the ratios between the intensity of the signal due to 
diffraction in the central pixel of our image should be the same as in the
case of a solar spectrum recorded with a similar video camera. Using such 
a recording and assuming that all the signal in the red channel is due 
to the Poisson spot, we obtained an excess of few percent in the blue and
white lights.

Once the residual flux of photons corresponding to the central pixel $\Phi_
\gamma$ is determined, the lifetime $\tau$ of the $\nu_2$ neutrino with 
respect to its radiative decay (in the earth reference system) could be 
computed from:

\begin{equation}
\Phi_\gamma = \epsilon \Phi_\nu \sin^2 \theta \left( 1 - e^{\frac
{t_{M \rightarrow E}}{\tau}} \right) e^{\frac{t_{S \rightarrow M}}{\tau}}.
\end{equation}
Here $\epsilon$ is the fraction of visible photons produced trough the 
radiative decay of neutrinos, $\Phi_\nu \sin^2 \theta$ represents the 
component $\nu_2$ of the expected solar neutrino flux at the earth location;
$t_{S \rightarrow M}$  and $t_{M \rightarrow E}$ are the average times of flight
of the $\nu_2$ neutrinos from the Sun to the Moon and from the Moon to the 
Earth, respectively.

By solving Eq. 4 for different choices of the oscillation parameters (as
shown in Fig: 1) we obtained limits on the $\nu_2$ lifetime of the order of
$10^6$s (in the earth reference frame). 

As already stressed, our indirect calibration has large uncertainties 
(estimated to be at the level of 25\% in terms of the 
photon flux); thus the above
result represents more an estimate of the achieved sensitivity than 
an actual limit.

\section{ The June 21, 2001, total solar eclipse}

The total solar eclipse of June 21, 2001, that crossed  Southern Africa,
allowed us to repeat the experiment. This time we 
performed only ground observations,
from a location $14^\circ 56'$ lat. S, 
$28^\circ 14'$ long. E. 
This location was at an altitude of about 1200 m a.s.l., 
at approximately 8 km from the 
central line of totality, and 
at about 50 km North of Lusaka, Zambia. From this location we could observe 
the totality for about 3.5 minutes. 

 We used two small Matsukov Cassagrain telescopes and a digital 
video-camera.

a) One of the Cassagrain telescopes (with a 12.5 cm aperture) was equipped
with a web-camera (and a small TV camera for pointing purposes). 
The recorded data 
consist in a digital film of the central part of the
dark side of the Moon (about 3000 single frames, each with an exposure time
of 1/25 s) 

b) The second Cassagrain (9 cm of aperture) had a digital photo-camera that
recorded 12 high
resolution (1600 $\times$ 1280 pixels$^2$) digital photographs, integrating
24~s of exposure. We took care to slightly change each time the field of view
in order to prevent some possible spurious reflection effects.

c) The digital video-camera was equipped with a ($2\times$) telelens and a
$10 \times$ optical zoom; 
it  produced a digital
recording equivalent to about 8000 single frames. 

The quality of all images seems good and we should have avoided
  the drawbacks reported
in Section 3. 
The data obtained by the three methods are currently under analysis; 
preliminary results seem to support the conclusions from the 1999 eclipse.

\section{Conclusions}

During the 1999 and 2001 total solar eclipses we looked for possible 
solar neutrino decays, which lead to an optical signature (visible light).

The analysis of the data extracted from the television recording of the 1999 
total solar 
eclipse is essentially completed. Some conclusions could be drawn.

a) An important surce of background was identified as the moonscape
seen in the light reflected by the Earth.

b) After removing the moon image and applying a wavelet decomposition
to the data, a peak remained, in all the color channels at a small 
angular distance from the center of the Moon (Sun); the peak could be produced
by the diffraction mechanism known as the ``Poisson spot"; also with this 
assumption we could not completely remove the signal.

c) 
The estimated sensitivity of our analysis in terms of the $\nu_2$ lifetime
is about $10^6$s (in the earth reference system).

The 2001 data were collected in good meteorological conditions, with three
different instruments with different  
resolutions, duty cicles and magnifications.  Preliminary 
analyses of  subsets of the data 
indicate much better quality data.
 The analyses of the bulk of the data and the calibrations of the 
equipments are in progress. We expect to obtain an improvement of at least
one order of magnitude with respect to the 1999 results. 

\section{Acknowledgements}

We acknowledge the National Institute for Aero-Space Research, Bucharest,
Romania and the Romanian Air Force for their co-operation in preparing 
the 1999 airborne experiment. We thank V\^{a}lcea 1 Television of 
R\^{a}mnicu V\^{a}lcea, Romania, for providing us with their video record 
of the eclipse.

The 2001 expedition in Zambia was financed by the Italian Space Agency (ASI) 
and by INFN. We are grateful to Kiboko Safaris, Malawi, for the efforts 
to offer us the best observation conditions possible. 

This work was partially supported by 
NATO grants CRG.LG. 972840, CN.SUPP 974683 and PST.CLG. 977691.

\end{document}